\documentstyle[amssymb,aps,12pt]{revtex}

\begin{document}
\draft
\author{Sergio De Filippo\cite{byline}}
\address{Dipartimento di Fisica ''E. R. Caianiello'', Universit\`{a} di Salerno\\
Via Allende I-84081 Baronissi (SA) ITALY\\
Tel: +39 089 965 229, Fax: +39 089 965 275, e-mail: defilippo@sa.infn.it\\
and \\
Unit\`{a} INFM Salerno}
\date{\today}
\title{Nonrelativistic field theoretic setting for gravitational self-interactions.}
\maketitle

\begin{abstract}
It is shown that a recently proposed model for the gravitational interaction
in non relativistic quantum mechanics is the instantaneous action at a
distance limit of a field theoretic model containing a negative energy
field. It reduces to the Schroedinger-Newton theory in a suitable mean field
approximation. While both the exact model and its approximation lead to
estimates for localization lengths, only the former gives rise to an
explicit non unitary dynamics accounting for the emergence of the classical
behavior of macroscopic bodies.
\end{abstract}


In a recent paper\cite{defilippo} a model for the gravitational interaction
in nonrelativistic quantum mechanics was proposed. In it matter degrees of
freedom were duplicated and gravitational interactions were introduced
between observable and unobservable degrees of freedom only. The non unitary
dynamics one is led to, once unobservable degrees of freedom are traced out,
includes both the traditional aspects of classical gravitational
interactions and a form of fundamental decoherence, which may be connected
with the emergence of the classical behavior of macroscopic bodies. The
model in fact treats on an equal footing mutual and self-interactions,
which, by some authors, are possibly held responsible for wave function
localization and/or reduction\cite{karolyhazy,diosi0,diosi,penrose,kumar}.

Interactions between observable and unobservable degrees of freedom are
instantaneous action at a distance ones and at first sight it looks unlikely
that they can be obtained as a nonrelativistic limit of more familiar local
interactions mediated by quantized fields. If, in fact, a local interaction
were introduced between an ordinary field and observable and unobservable
matter, the corresponding low energy limit would include an action at a
distance inside the observable (and the unobservable) matter too. In this
letter we want to show that a field theoretic reading of the model emerges
naturally through a Stratonovich-Hubbard transformation\cite{negele} of the
gravitational interaction. Within the minimal possible generalization of the
model, where the instantaneous interaction is replaced by a retarded
potential, the result of the transformation corresponds to the emergence of
an ordinary scalar field and a negative energy one, both coupled with the
matter through a Yukawa interaction. This result has a plain physical
reading, as it gives rise to two competing interactions with overall
vanishing effect within observable (and unobservable) matter: an attractive
and a repulsive one respectively mediated by the positive and the negative
energy field.

The presence of a negative energy field somehow is not surprising if we
consider that, in a dynamical theory supposed to account for wave function
localization, one expects a small continuous energy injection\cite
{squires,ring}, and negative energy fields are the most natural candidates
for that, apart from the possible introduction of a cosmological background.
In fact the possible role of negative-energy fields was already suggested
within some attempts to account for wave function collapse by
phenomenological stochastic models\cite{pearle}.

As a by-product of the Stratonovich-Hubbard transformation we show also that
a proper mean field approximation leads to the Schroedinger-Newton (S-N)
model \cite{kumar}. Finally, while the original model and the S-N
approximation are equivalent as to the classical aspects of the
gravitational interaction, the S-N model is shown to be ineffective in
turning off quantum coherences corresponding to different locations of one
and the same macroscopic body, at variance with the original model.

To be specific, following Ref. \cite{defilippo}, let $H[\psi ^{\dagger
},\psi ]$ denote the second quantized non-relativistic Hamiltonian of a
finite number of particle species, like electrons, nuclei, ions, atoms
and/or molecules, according to the energy scale. For notational simplicity $%
\psi ^{\dagger },\psi $ denote the whole set $\psi _{j}^{\dagger }(x),\psi
_{j}(x)$ of creation-annihilation operators, i.e. one couple per particle
species and spin component. This Hamiltonian includes the usual
electromagnetic interactions accounted for in atomic and molecular physics.
To incorporate gravitational interactions including self-interactions, we
introduce complementary creation-annihilation operators $\chi _{j}^{\dagger
}(x),\chi _{j}(x)$ and the overall Hamiltonian 
\begin{equation}
H_{G}=H[\psi ^{\dagger },\psi ]+H[\chi ^{\dagger },\chi
]-G\sum_{j,k}m_{j}m_{k}\int dxdy\frac{\psi _{j}^{\dagger }(x)\psi
_{j}(x)\chi _{k}^{\dagger }(y)\chi _{k}(y)}{|x-y|},  \label{hamiltonian}
\end{equation}
acting on the tensor product $F_{\psi }\otimes F_{\chi }$ of the Fock spaces
of the $\psi $ and $\chi $ operators, where $m_{i}$ denotes the mass of the $%
i$-th particle species and $G$ is the gravitational constant. While the $%
\chi $ operators are taken to obey the same statistics as the original
operators $\psi $, we take advantage of the arbitrariness pertaining to
distinct operators and, for simplicity, we choose them commuting with one
another: $[\psi ,\chi ]$ $_{-}=[\psi ,\chi ^{\dagger }]_{-}=0$.

The metaparticle state space $S$ is identified with the subspace of $F_{\psi
}\otimes F_{\chi }$ including the metastates obtained from the vacuum $%
\left| 0\right\rangle =\left| 0\right\rangle _{\psi }\otimes \left|
0\right\rangle _{\chi }$ by applying operators built in terms of the
products $\psi _{j}^{\dagger }(x)\chi _{j}^{\dagger }(y)$ and symmetrical
with respect to the interchange $\psi ^{\dagger }\leftrightarrow \chi
^{\dagger }$, which, as a consequence, have the same number of $\psi $
(green) and $\chi $ (red) metaparticles of each species. This is a
consistent definition since the overall Hamiltonian is such that the
corresponding time evolution is a group of (unitary) endomorphisms of $S$.
If we prepare a pure $n$-particle state, represented in the original setting
- excluding gravitational interactions - by 
\begin{equation}
\left| g\right\rangle \doteq \int d^{n}xg(x_{1},x_{2},...,x_{n})\psi
_{j_{1}}^{\dagger }(x_{1})\psi _{j_{2}}^{\dagger }(x_{2})...\psi
_{j_{n}}^{\dagger }(x_{n})\left| 0\right\rangle ,
\end{equation}
its representation in $S$ is given by the metastate 
\begin{equation}
\left| \left| g\otimes g\right\rangle \right\rangle =\int
d^{n}xd^{n}yg(x_{1},...,x_{n})g(y_{1},...,y_{n})\psi _{j_{1}}^{\dagger
}(x_{1})...\psi _{j_{n}}^{\dagger }(x_{n})\chi _{j_{1}}^{\dagger
}(y_{1})...\chi _{j_{n}}^{\dagger }(y_{n})\left| 0\right\rangle .
\label{initial}
\end{equation}
As for the physical algebra, it is identified with the operator algebra of
say the green metaworld. In view of this, expectation values can be
evaluated by preliminarily tracing out the $\chi $ operators and then taking
the average in accordance with the traditional setting.

While we are talking trivialities as to an initial metastate like in Eq. (%
\ref{initial}), that is not the case in the course of time, since the
overall Hamiltonian produces entanglement between the two metaworlds,
leading, once $\chi $ operators are traced out, to mixed states of the
physical algebra. It was shown in Ref. \cite{defilippo} that the ensuing
non-unitary evolution induces both an effective interaction mimicking
gravitation, and wave function localization. Localization is kept in time,
as the spreading of the probability density of the center of mass of a
macroscopic body does not imply a spreading of the wave function, but rather
it is due to the emergence of a delocalized ensemble of localized pure
states \cite{defilippo1}. A peculiar feature of the model is that it cannot
be obtained by quantizing its naive classical version, since the classical
states corresponding to the constraint in $F_{\psi }\otimes F_{\chi }$,
selecting the metastate space $S$, have green and red partners sitting in
the same space point and then a divergent gravitational energy. While it is
usual that, in passing from the classical to the quantum description,
self-energy divergences are mitigated, in this instance we pass from a
completely meaningless classical theory to a quite divergence free one. This
is more transparent below, where we consider the field theoretic description.

Let us adopt here an interaction representation, where the free Hamiltonian
is identified with $H[\psi ^{\dagger },\psi ]+H[\chi ^{\dagger },\chi ]$ and
the time evolution of an initially untangled metastate $\left| \left| \tilde{%
\Phi}(0)\right\rangle \right\rangle $\ is represented by 
\begin{eqnarray}
\left| \left| \tilde{\Phi}(t)\right\rangle \right\rangle &=&{\it T}\exp %
\left[ \frac{i}{\hslash }Gm^{2}\int dt\int dxdy\frac{\psi ^{\dagger
}(x,t)\psi (x,t)\chi ^{\dagger }(y,t)\chi (y,t)}{|x-y|}\right] \left| \left| 
\tilde{\Phi}(0)\right\rangle \right\rangle  \nonumber \\
&\equiv &U(t)\left| \left| \tilde{\Phi}(0)\right\rangle \right\rangle \equiv
U(t)\left| \Phi (0)\right\rangle _{\psi }\otimes \left| \Phi
(0)\right\rangle _{\chi }.  \label{evolvedmetastate}
\end{eqnarray}
Then, by making use of a Stratonovich-Hubbard transformation\cite{negele},
we can rewrite the time evolution operator in the form 
\begin{eqnarray}
U(t) &=&\int {\it D}\left[ \varphi _{1}\right] {\it D}\left[ \varphi _{2}%
\right] \exp \frac{ic^{2}}{2\hslash }\int dtdx\left[ \varphi _{1}\nabla
^{2}\varphi _{1}-\varphi _{2}\nabla ^{2}\varphi _{2}\right]  \nonumber \\
&&{\it T}\exp \left[ -i\frac{mc}{\hslash }\sqrt{2\pi G}\int dtdx\left[
\varphi _{1}(x,t)+\varphi _{2}(x,t)\right] \psi ^{\dagger }(x,t)\psi (x,t)%
\right]  \nonumber \\
&&{\it T}\exp \left[ -i\frac{mc}{\hslash }\sqrt{2\pi G}\int dtdx\left[
\varphi _{1}(x,t)-\varphi _{2}(x,t)\right] \chi ^{\dagger }(x,t)\chi (x,t)%
\right] ,  \label{stratonovich}
\end{eqnarray}
namely as a functional integral over two auxiliary real scalar fields $%
\varphi _{1}$ and $\varphi _{2}$.

To give a physical interpretation of this result, consider the minimal
variant of the Newton interaction in Eq. (\ref{evolvedmetastate}) aiming at
avoiding instantaneous action at a distance, namely consider replacing $%
-1/|x-y|$ by the Feynman propagator $4\pi \square ^{-1}\equiv 4\pi \left(
-\partial _{t}^{2}/c^{2}+\nabla ^{2}\right) ^{-1}$. Then the analog of Eq. (%
\ref{stratonovich}) holds with the d'alembertian $\square $ replacing the
Laplacian $\nabla ^{2}$ and the ensuing expression can be read as the mixed
path integral and operator expression for the evolution operator
corresponding to the field Hamiltonian 
\begin{eqnarray}
H_{Field} &=&H[\psi ^{\dagger },\psi ]+H[\chi ^{\dagger },\chi ]+\frac{1}{2}%
\int dx\left[ \pi _{1}^{2}+c^{2}\left| \nabla \varphi _{1}\right| ^{2}-\pi
_{2}^{2}-c^{2}\left| \nabla \varphi _{2}\right| ^{2}\right] 
\label{fieldhamiltonian} \\
&&+mc\sqrt{2\pi G}\int dx\left\{ \left[ \varphi _{1}+\varphi _{2}\right]
\psi ^{\dagger }\psi +\left[ \varphi _{1}-\varphi _{2}\right] \chi ^{\dagger
}\chi \right\} ,
\end{eqnarray}
where $\pi _{1}=\dot{\varphi}_{1}$ and $\pi _{2}=\dot{\varphi}_{2}$
respectively denote the conjugate fields of $\varphi _{1}$ and $\varphi _{2}$
and all fields denote quantum operators. This theory can be read in analogy
with nonrelativistic quantum electrodynamics, where a relativistic field is
coupled with nonrelativistic matter, while the procedure to obtain the
corresponding action at a distance theory by integrating out the $\varphi $
fields is the analog of the Feynman's elimination of electromagnetic field
variables\cite{feynman}.

The resulting theory, containing the negative energy field $\varphi _{2}$,
has the attractive feature of being divergence free, at least in the
non-relativistic limit, where Feynman graphs with virtual
particle-antiparticle pairs can be omitted. To be specific, it does not
require the infinite self-energy subtraction needed for instance in
electrodynamics on evaluating the Lamb shift, or the coupling constant
renormalization\cite{itzykson}. Here of course we refer to the covariant
perturbative formalism applied to our model, where matter fields are
replaced by their relativistic counterparts and the non-relativistic
character of the model is reflected in the mass density being considered as
a scalar coupled with the scalar fields by Yukawa-like interactions. In fact
there is a complete cancellation among all Feynman diagrams containing only $%
\psi $ (or equivalently $\chi $) and internal $\varphi $ lines, owing to the
difference in sign between the $\varphi _{1}$ and the $\varphi _{2}$ free
propagators. This state of affairs of course is the field theoretic
counterpart of the absence of direct $\psi -\psi $ and $\chi -\chi $
interactions in the theory obtained by integrating out the $\varphi $
operators, whose presence would otherwise require the infinite self-energy
subtraction corresponding to normal ordering. These considerations,
supported by the mentioned suggestions derived from a phenomenological
analysis\cite{pearle} about the possible role of negative energy fields, may
provide substantial clues for the possible extensions of the model towards a
relativistic theory of gravity-induced localization. On the other hand a
Yukawa interaction with a (positive energy) scalar field emerges also moving
from Einstein's theory of gravitation, if one confines consideration to
conformal space-time fluctuations in a linear approximation \cite
{rosales,power}.

Going back to our evolved metastate (\ref{evolvedmetastate}), the
corresponding physical state is given by 
\begin{equation}
M(t)\equiv Tr_{\chi }\left| \left| \tilde{\Phi}(t)\right\rangle
\right\rangle \left\langle \left\langle \tilde{\Phi}(t)\right| \right|
=\sum_{k}\;\;\;_{\chi }\left\langle k\right| \left| \left| \tilde{\Phi}%
(t)\right\rangle \right\rangle \left\langle \left\langle \tilde{\Phi}%
(t)\right| \right| \left| k\right\rangle _{\chi },
\end{equation}
and, by using Eq. (\ref{stratonovich}), we can write 
\begin{eqnarray}
_{\chi }\left\langle k\right| \left| \left| \tilde{\Phi}(t)\right\rangle
\right\rangle  &=&\int {\it D}\left[ \varphi _{1}\right] {\it D}\left[
\varphi _{2}\right] \exp \frac{ic^{2}}{2\hslash }\int dtdx\left[ \varphi
_{1}\nabla ^{2}\varphi _{1}-\varphi _{2}\nabla ^{2}\varphi _{2}\right]  
\nonumber \\
&&_{\chi }\left\langle k\right| {\it T}\exp \left[ -i\frac{mc}{\hslash }%
\sqrt{2\pi G}\int dtdx\left[ \varphi _{1}(x,t)-\varphi _{2}(x,t)\right] \chi
^{\dagger }(x,t)\chi (x,t)\right] \left| \Phi (0)\right\rangle _{\chi } 
\nonumber \\
&&{\it T}\exp \left[ -i\frac{mc}{\hslash }\sqrt{2\pi G}\int dtdx\left[
\varphi _{1}(x,t)+\varphi _{2}(x,t)\right] \psi ^{\dagger }(x,t)\psi (x,t)%
\right] \left| \Phi (0)\right\rangle _{\psi }.
\end{eqnarray}
Then the final expression for the physical state at time $t$ is given by 
\begin{equation}
M(t)=  \label{exact}
\end{equation}
\begin{eqnarray*}
&&\int {\it D}\left[ \varphi _{1}\right] {\it D}\left[ \varphi _{2}\right] 
{\it D}\left[ \varphi _{1}^{\prime }\right] {\it D}\left[ \varphi
_{2}^{\prime }\right] \exp \frac{ic^{2}}{2\hslash }\int dtdx\left[ \varphi
_{1}\nabla ^{2}\varphi _{1}-\varphi _{2}\nabla ^{2}\varphi _{2}-\varphi
_{1}^{\prime }\nabla ^{2}\varphi _{1}^{\prime }+\varphi _{2}^{\prime }\nabla
^{2}\varphi _{2}^{\prime }\right]  \\
&&_{\chi }\left\langle \Phi (0)\right| {\it T}^{-1}\exp \left[ i\frac{mc}{%
\hslash }\sqrt{2\pi G}\int dtdx\left[ \varphi _{1}^{\prime }-\varphi
_{2}^{\prime }\right] \chi ^{\dagger }\chi \right] {\it T}\exp \left[ -i%
\frac{mc}{\hslash }\sqrt{2\pi G}\int dtdx\left[ \varphi _{1}-\varphi _{2}%
\right] \chi ^{\dagger }\chi \right] \left| \Phi (0)\right\rangle _{\chi } \\
&&{\it T}\exp \left[ -i\frac{mc}{\hslash }\sqrt{2\pi G}\int dtdx\left[
\varphi _{1}+\varphi _{2}\right] \psi ^{\dagger }\psi \right] \left| \Phi
(0)\right\rangle _{\psi \psi }\left\langle \Phi (0)\right| {\it T}^{-1}\exp i%
\frac{mc}{\hslash }\sqrt{2\pi G}\int dtdx\left[ \varphi _{1}^{\prime
}+\varphi _{2}^{\prime }\right] \psi ^{\dagger }\psi ,
\end{eqnarray*}
where, due to the constraint on the metastate space, $\chi $ operators can
be replaced by $\psi $ operators, if simultaneously the metastate vector $%
\left| \Phi (0)\right\rangle _{\chi }$ is replaced by $\left| \Phi
(0)\right\rangle _{\psi }$. Then, if in the c-number factor corresponding to
the $\chi $-trace, we make the mean field (MF) approximation $\psi ^{\dagger
}\psi \rightarrow \left\langle \psi ^{\dagger }\psi \right\rangle $, we get 
\[
M_{MF}(t)=
\]
\begin{eqnarray}
&&\int {\it D}\left[ \varphi _{1}\right] {\it D}\left[ \varphi _{2}\right] 
{\it D}\left[ \varphi _{1}^{\prime }\right] {\it D}\left[ \varphi
_{2}^{\prime }\right] \exp \frac{ic^{2}}{2\hslash }\int dtdx\left[ \varphi
_{1}\nabla ^{2}\varphi _{1}-\varphi _{2}\nabla ^{2}\varphi _{2}-\varphi
_{1}^{\prime }\nabla ^{2}\varphi _{1}^{\prime }+\varphi _{2}^{\prime }\nabla
^{2}\varphi _{2}^{\prime }\right]   \nonumber \\
&&{\it T}\exp \left[ -i\frac{mc}{\hslash }\sqrt{2\pi G}\int dtdx\left[
\varphi _{1}\left[ \psi ^{\dagger }\psi +\left\langle \psi ^{\dagger }\psi
\right\rangle \right] +\varphi _{2}\left[ \psi ^{\dagger }\psi -\left\langle
\psi ^{\dagger }\psi \right\rangle \right] \right] \right] \left| \Phi
(0)\right\rangle _{\psi }  \nonumber \\
&&_{\psi }\left\langle \Phi (0)\right| {\it T}^{-1}\exp i\frac{mc}{\hslash }%
\sqrt{2\pi G}\int dtdx\left[ \varphi _{1}^{\prime }\left[ \psi ^{\dagger
}\psi +\left\langle \psi ^{\dagger }\psi \right\rangle \right] +\varphi
_{2}^{\prime }\left[ \psi ^{\dagger }\psi -\left\langle \psi ^{\dagger }\psi
\right\rangle \right] \right] .
\end{eqnarray}
Finally, if we perform functional integrations, we get 
\begin{eqnarray}
M_{MF}(t) &=&  \nonumber \\
&&{\it T}\exp \left[ \frac{i}{\hslash }Gm^{2}\int dt\int dxdy\frac{\psi
^{\dagger }(x,t)\psi (x,t)\left\langle \psi ^{\dagger }(y,t)\psi
(y,t)\right\rangle }{|x-y|}\right] \left| \Phi (0)\right\rangle _{\psi } 
\nonumber \\
&&_{\psi }\left\langle \Phi (0)\right| {\it T}^{-1}\exp \left[ -\frac{i}{%
\hslash }Gm^{2}\int dt\int dxdy\frac{\psi ^{\dagger }(x,t)\psi
(x,t)\left\langle \psi ^{\dagger }(y,t)\psi (y,t)\right\rangle }{|x-y|}%
\right] ,  \label{meanfield}
\end{eqnarray}
namely in this approximation the model is equivalent to the S-N theory\cite
{kumar}. Of course the whole procedure could be repeated without substantial
variations starting from the field theoretic Hamiltonian (\ref
{fieldhamiltonian}), inserting the mean field approximation before applying
the Feynman's procedure for the elimination of field variables, and then
getting a retarded potential version of the S-N model.

However this approximation does not share with the original model the
crucial ability of making linear superpositions of macroscopically different
states unobservable. Consider in fact an initial state corresponding to the
linear, for simplicity orthogonal, superposition of $N$ localized states of
one and the same macroscopic body, which were shown to exist as pure states
corresponding to unentangled bound metastates of green and red metamatter
for bodies of ordinary density and a mass higher than $\sim 10^{11}$ proton
masses\cite{defilippo}: 
\begin{equation}
\left| \Phi (0)\right\rangle =\frac{1}{\sqrt{N}}\sum_{j=1}^{N}\left|
z_{j}\right\rangle ,  \label{superposition}
\end{equation}
where $\left| z\right\rangle $ represents a localized state centered in $z$.
Compare the coherence $\left\langle z_{h}\right| M(t)\left|
z_{k}\right\rangle $ when evaluated according to Eqs. (\ref{exact}) and (\ref
{meanfield}), where we consider the localized states as approximate
eigenstates of the particle density operator 
\begin{equation}
\psi ^{\dagger }(x,t)\psi (x,t)\left| z\right\rangle \simeq n(x-z)\left|
z\right\rangle
\end{equation}
and time dependence in $\psi ^{\dagger }\psi $ irrelevant, as the considered
states are stationary states in the Schroedinger picture apart from an
extremely slow spreading\cite{defilippo1}.

According to the original model we get, from Eq. (\ref{exact}) 
\begin{eqnarray}
&&\left\langle z_{h}\right| M(t)\left| z_{k}\right\rangle  \nonumber \\
&=&\int {\it D}\left[ \varphi _{1}\right] {\it D}\left[ \varphi _{2}\right] 
{\it D}\left[ \varphi _{1}^{\prime }\right] {\it D}\left[ \varphi
_{2}^{\prime }\right] \exp \frac{ic^{2}}{2\hslash }\int dtdx\left[ \varphi
_{1}\nabla ^{2}\varphi _{1}-\varphi _{2}\nabla ^{2}\varphi _{2}-\varphi
_{1}^{\prime }\nabla ^{2}\varphi _{1}^{\prime }+\varphi _{2}^{\prime }\nabla
^{2}\varphi _{2}^{\prime }\right]  \nonumber \\
&&\frac{1}{N^{2}}\sum_{j=1}^{N}\exp \left[ -i\frac{mc}{\hslash }\sqrt{2\pi G}%
\int dtdx\left[ \left[ \varphi _{1}-\varphi _{2}\right] n(x-z_{j})-\left[
\varphi _{1}^{\prime }-\varphi _{2}^{\prime }\right] n(x-z_{j})\right] %
\right]  \nonumber \\
&&\exp \left[ -i\frac{mc}{\hslash }\sqrt{2\pi G}\int dtdx\left[ \left[
\varphi _{1}+\varphi _{2}\right] n(x-z_{h})-\left[ \varphi _{1}^{\prime
}+\varphi _{2}^{\prime }\right] n(x-z_{k})\right] \right] ,
\end{eqnarray}
and, after integrating out the scalar fields, 
\begin{equation}
\left\langle z_{h}\right| M(t)\left| z_{k}\right\rangle =\frac{1}{N^{2}}%
\sum_{j=1}^{N}\exp \frac{i}{\hslash }Gm^{2}t\int dxdy\left[ \frac{%
n(x-z_{j})n(y-z_{h})}{|x-y|}-\frac{n(x-z_{j})n(y-z_{k})}{|x-y|}\right] ,
\label{coherences}
\end{equation}
which shows that, while diagonal coherences are given by $\left\langle
z_{h}\right| M(t)\left| z_{h}\right\rangle =1/N$, the off-diagonal ones,
under reasonable assumptions on the linear superposition in Eq. (\ref
{superposition}) of a large number of localized states, approximately
vanish, due to the random phases in the sum in Eq. (\ref{coherences}). This
makes the state $M(t)$, for not too short times, equivalent to an ensemble
of localized states: 
\begin{equation}
M(t)\simeq \frac{1}{N}\sum_{j=1}^{N}\left| z_{j}\right\rangle \left\langle
z_{j}\right| .
\end{equation}
On the other hand, if we calculate coherences according to the S-N model, we
get 
\begin{eqnarray}
&&\left\langle z_{h}\right| M_{MF}(t)\left| z_{k}\right\rangle  \nonumber \\
&=&\frac{1}{N}\exp \frac{i}{\hslash }Gm^{2}t\int dxdy\frac{\left[
n(x-z_{h})-n(x-z_{k})\right] \sum_{j=1}^{N}n(y-z_{j})/N}{|x-y|},
\end{eqnarray}
so that here the sum appears in the exponent and there is no cancellation.
While diagonal coherences keep the same value as in the original model,
off-diagonal ones acquire only a phase for the presence of the mean
gravitational interaction, but keep the same absolute value $1/N$ as the
diagonal ones. Furthermore, if we take $N=2$ rather than $N$ very large, the
S-N approximation gives just $\left\langle z_{1}\right| M_{MF}(t)\left|
z_{2}\right\rangle =1/2$, whereas the exact model still offers a mechanism
to make off-diagonal coherences unobservable, due to time oscillations\cite
{defilippo}.

It is worth while to remind that, although both our proposal and the S-N
model give rise to localized states and reproduce the classical aspects of
the gravitational interaction, the mean field approximation, necessary to
pass from the former to the latter, spoils the theory, not only of its
feature of reducing unlocalized wave functions, but also of another
desirable property. In fact it can be shown that according to our model
localized states evolve into unlocalized ensembles of localized states\cite
{defilippo1}, while the S-N theory leads to stationary localized states\cite
{kumar}, which is rather counterintuitive and unphysical, since space
localization implies linear momentum uncertainty, and this, in its turn,
should imply a spreading of the probability distribution in space.

In conclusion, while the present model has only a nonrelativistic character,
its analysis hints of possible directions for extensions to higher energies,
where an instantaneous action at a distance is not appropriate. In
particular, the emergence of negative energy fields leads naturally to a
promising perspective for the construction of finite field theories, where
divergence cancellations are due to the presence of couples of positive and
negative energy fields, rather than of supersymmetric partners. Furthermore,
since the geometric formulation of Newtonian gravity, i.e. the Newton-Cartan
theory, leads only to the mean field approximation, i.e. the S-N theory\cite
{christian}, it may be likely that the geometric aspects of gravity may even
play a misleading role in looking for a quantum theory including gravity
both in its classical aspects and in its possible localization effects. More
specifically the Einstein theory of gravitation could arise, unlike, for
instance, classical electrodynamics, not as a result of taking expectation
values with respect to a pure physical state, but rather, as an effective
long distance theory like hydrodynamics, from a statistical average, or
equivalently by tracing out unobservable degrees of freedom starting from a
pure metastate.

Acknowledgments - Financial support from M.U.R.S.T., Italy and I.N.F.M.,
Salerno is acknowledged

\end{document}